\documentclass[fleqn,twoside]{article}
\usepackage{espcrc2}

\usepackage{graphicx}

\newcommand{\AmS}{{\protect\the\textfont2
  A\kern-.1667em\lower.5ex\hbox{M}\kern-.125emS}}

\newcommand{\be}{\begin{equation}}
\newcommand{\ee}{\end{equation}}

\newcommand{\bea}{\begin{eqnarray*}}
\newcommand{\eea}{\end{eqnarray*}}

\newcommand{\pspic}[4]{
\begin{center}
\begin{tabular}{c}
\includegraphics[angle=270,width={#2}]{#1}
\end{tabular}
\end{center}
\caption{#3}
\label{#4}
\end{figure}
}

\title{A Test of The Source Galerkin Method}

\author{D. Petrov\address[HET]{Physics Department,
        Brown University, \\
        Providence, RI 02912},
        P. Emirdag\addressmark[HET],
        G. S. Guralnik\addressmark[HET]}

\begin{document}

\begin{abstract}
Some results of the ongoing development of our Source Galerkin (SG)
nonperturbative approach to numerically solving Quantum Field theories
are presented.  This technique has the potential to be much faster
than Monte Carlo methods. SG uses known symmetries and theoretical
properties of a theory. In order to test this approach, we applied it
to $\phi^4$ theory in zero dimensions. This model has been extensively
studied and has a known set of exact solutions. This
allows us to broaden the understanding of various properties of the SG
method and to develop techniques necessary for the successful application
of this method to more sophisticated theories.
\end{abstract}

\maketitle

\section{Introduction.}
The Source Galerkin method is being developed as a flexible alternative to Monte
Carlo approaches to solving Quantum Field Theories.

\subsection{Overview of the Method}
To illustrate the application of the Source Galerkin approach to solving Quantum

Field Theories we consider a theory defined by a Lagrangian $\cal L$. All
information of the theory is contained in the generating functional
\be
        Z(j) = \int {\cal D}\phi \exp\left[\int(-{\cal L} + j\phi)d^dx\right].
\ee
This satisfies a Schwinger-Dyson set of differential equations.
A procedure to determine an approximate solution of 
these equations is specified by our SG technique. To start, a trial 
solution or an ansatz must be introduced. In general, it is constructed from a
predefined set of trial functions. The proper choice of trial functions
allows us to take advantage of symmetries of the theory as well as of other 
known analytical properties of the generating functional. A simple example of an
approximate solution is:
\be
        Z_a = \exp\left[\int j_x G_{xy} j_y +\right.
         \left. \int j_wj_xH_{wxyz}j_yj_z + ... \right].
\ee
A set of residues is obtained by substituting this ansatz into the
Schwinger-Dyson equations. The undefined coefficients of the trial solution are
determined by solving a system of nonlinear equations obtained by projecting the
residues on the trial functions and requiring these projections to be equal to
zero.
The procedure outlined above guarantees that the error associated with the
approximate solution converges to zero in the mean as the number of members in 
the 
set of trial functions goes to infinity.

\subsection{Motivation}
In order to demonstrate the validity of the method, it was applied to the $O(3)$
nonlinear $\sigma$ model \cite{Emirdag99}. This model is asymptotically free and
is a useful toy model for approaching Non-Abelian gauge theories. First order 
solutions were obtained, but during our attempts to extend this work 
to higher orders it became evident that a better understanding of the general 
properties of the SG method is necessary. Theorems guarantee that 
Galerkin methods produce an approximate solution, which converges to the exact 
solution as
the number of terms goes to infinity. In practice, the approximate solution can 
only have a fairly small number of parameters. Therefore, we must 
ensure that high accuracy and rapid convergence can reasonably
be achieved with only a few 
terms in the ansatz. The fact that the error approaches zero in the mean
implies that the accuracy of the result is heavily affected by the choice of 
inner product.
We attempt to investigate the effectiveness of the Source Galerkin approach by
using it to solve $\phi^4$ theory in zero dimensions. We study its performance 
for several choices of trial functions and scalar products.

\section {Ultralocal Model}
After introducing an external source $j$, we define the ultralocal model by
the following lagrangian
\be
        {\cal L} = \frac{g}{4}\phi^4 + \frac{\mu}{2}\phi^2 - j\phi.
\ee
Theoretical solutions for this theory can be obtained by expressing generating 
functional as a power series
\be
        Z_a(j) = \sum\limits_{k=0}^{\infty}\frac{1}{k!}a_kj^k.
\ee
The coefficients of this series are given by
\cite{Garcia96}
\begin{small}
\bea
	a_{2n}&=&(2n-1)!!\frac{U(n,t)+(-1)^n\rho U(n,-t)}{U(0,t)+\rho U(0,-t)},\\
	a_{2n+1}&=&\frac{2n!!}{n!}(-t)^n\frac{V(n+\frac{1}{2},t)}{V(\frac{1}{2},t)}
	\alpha\frac{
	t^{\frac{1}{2}}e^{\frac{t^2}{4}} }{U(0,t)+\rho U(0,-t)}.
\eea
\end{small}
Here $U$ and $V$ are parabolic cylinder functions, coefficients $\rho$ and
$\alpha$ fix boundary conditions.

\subsection {Numerical Solution}
In analogy with the theoretical solution, a truncated power series of order $N$
can be chosen as a trial solution.
After obtaining the residues  by substituting this expression
into the Schwinger-Dyson equations, we need to eliminate source dependence by
projecting them on the trial functions. Two possible ways to define scalar 
product are presented below.
\begin{small}
\be
        \int\limits^c_{-c}j^nR(j)dj,\;\;\;\; n = 0..N-3
\ee
and
\be
        \int\limits^\infty_{-\infty}j^nR(j)\exp{
        \left[{-\frac{j^2}{\epsilon^2}}\right]}dj,\;\;\;\; n = 0..N-3.
\ee
\end{small}
Both definitions are localized in the region close to $j=0$ since eventually we
want to set the source to zero in order to compute physical values. Using these
definitions produces similar results as shown in figure 1. However, the
second expression is more general and it is easier to extend to calculations 
with
space-time dimensions.
\begin{figure}[t]
\includegraphics[angle=270,width=8cm]{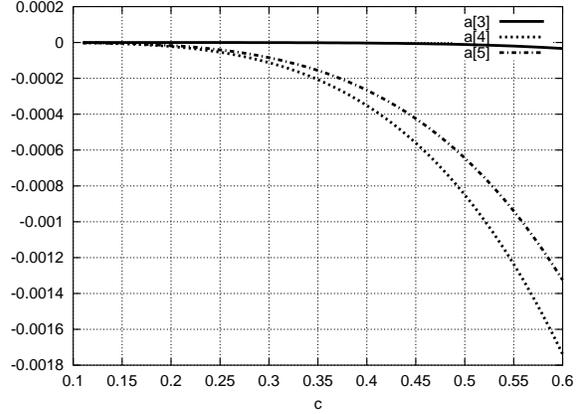}
\caption{ Order by order relative error is plotted for N=5 and residual
equations derived from {$\int\limits^c_{-c}j^nR(j)dj$} with respect to the
range of integration $c$}
\label{fig1}
\end{figure}
\begin{figure}[t]
\includegraphics[angle=270,width=8cm]{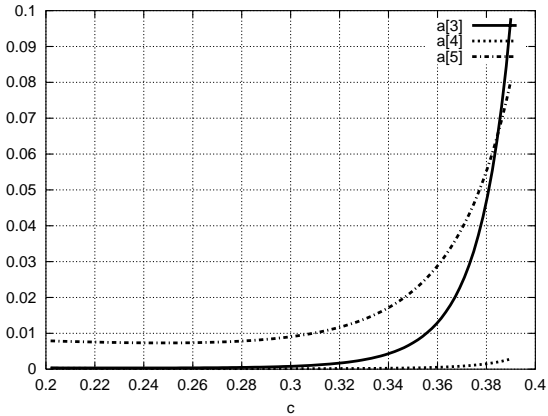}
\caption{Order by order relative error is plotted
for $N=7$ and $N=5$. Residual equations derived from
$\int\limits^\infty_{-\infty}H_n(j)R(j)\exp{\left[-\frac{j^2}{\epsilon^2}\right]}dj$
with respect to the parameter $\epsilon$.}
\label{fig1}
\end{figure}
Inspection of these results shows that the error in determination of
coefficients $a_i$ increases rapidly at high orders.
In spite of this, the generating functional can
be determined with accuracy as high as several parts in $10^7$ for a significant
range of j. However, it must be noted that correct determination of
the proper range
of integration or a value of the parameter $\epsilon$ is crucial for achieving 
high
accuracy. The SG approach by itself does not provide an algorithm for 
setting these parameters. In a real problem when the exact result is
not available it is necessary to have some external source of information which
allows us to determine the correct values of the parameters.

\subsection{Solution in Terms of Hermite Polynomials}
A set of Hermite polynomials can be defined by
\begin{small}
\be
        H_n(\xi)= (-1)^n \exp{\left\{\frac{\xi ^2}{\epsilon^2}\right\}}
\frac{d^n
                \exp{\left\{-\frac{\xi ^2}{\epsilon^2}\right\}}}{d \xi^n}.
\ee
\end{small}
These polynomials are orthonormal under the following inner product
\begin{small}
\be
        \int\limits_{-\infty}^{\infty} H_k(x)
        H_l(x)\exp{\left(-\frac{x^2}{\epsilon^2}\right)}
                 dx = \delta^{kl}.
\ee
\end{small}
This suggests a way to improve our implementation of the Source Galerkin 
procedure for the ultralocal theory. Trial functions can be constructed as a
linear combination of Hermite polynomials of different orders, while the
expression
shown above is used to define the scalar product. This choice leads to 
significant 
simplification of the system of equations which give the coefficients in the 
ansatz. 
Figure 2 shows the dependence of relative error in determination of values $a_i$
on the parameter $\epsilon$. From this graph we observe that there is a 
significant range of values of the parameters in which numerical results remain 
stable. We can use this fact to resolve the problem outlined in the end of
previous section. In a real problem, it would be possible to set all the 
parameters without any knowledge of theoretical solution just by finding
solutions for all possible values of parameters and then choosing a solution 
which is stable within the required accuracy for some range of the parameters.
The accuracy of determination of the generating functional using this method is
consistent with the power series ansatz approach. This implies that it is 
limited only by error introduced by the method used to solve 
the system of equations produced by the Source Galerkin procedure and
by the numerical precision of the hardware used for computation.

\section{Conclusion}
This investigation demonstrates that very high precision computations can be 
performed using the Source Galerkin procedure. On the other hand, the method is 
very sensitive to the choice of parameters. Application of this method
to a more 
general set of problems will require development of some consistent method of 
determination of proper values of parameters. Current results suggest that use 
of orthogonal trial functions may simplify this task.
Source Galerkin technique is still under development. Additional work is needed 
to fully investigate it's properties and to apply it to various field theories.

\section{Acknowledgements:} We wish to thank R. Easther, D. Ferrante,
Z. Guralnik and S. Hahn for helpful discussions. This research was
supported in part by DOE grant DOE grant DE-FG02/19ER40688-(Task D).

\end{document}